\documentclass{JHEP3}

\usepackage{amsmath}
\input epsf
\usepackage{epsfig}
\usepackage{amssymb}
\usepackage{graphics}
\usepackage[active]{srcltx}

\newcommand{\bea}{\begin{eqnarray}}
\newcommand{\eea}{\end{eqnarray}}
\newcommand{\bean}{\begin{eqnarray*}}
\newcommand{\eean}{\end{eqnarray*}}
\newcommand{\nn}{\nonumber \\}

\def\W #1{\widetilde{#1}}
\def\WH #1{\widehat{#1}}

\def\eref#1{(\ref{#1})}

\def\a{{\alpha}}

\def\b{{\beta}}

\def\Sl{\sum\limits}
\def\Label#1{\label{#1}%
  \smash{\hbox to0pt{\raise1ex\hbox{\tiny[#1]}\hss}}}

\allowdisplaybreaks
\title{Note on symmetric BCJ numerator}

\author{ Chih-Hao Fu ${}^a$\footnote{The unusual ordering of authors in this paper is just
 to fit outdated requirement of Chinese University for recognition of contributions.},  Yi-Jian Du ${}^{b,c}$, Bo Feng ${}^{d,e}$~~~~~~~~~~~~~~\\
${}^a$ Department of Electrophysics, National Chiao Tung University\\
1001 University Road, Hsinchu, Taiwan, R.O.C.
\\${}^b$ Department of Astronomy and Theoretical Physics, Lund University
SE-22362 Lund, Sweden
\\${}^c$ Department of Physics and Center for Field Theory
and Particle Physics, Fudan University, 220 Handan Road, Shanghai 200433, P.R China
\\${}^d$ Zhejiang Institute of Modern Physics, Zhejiang
University, 38 Zheda Road Hangzhou, 310027 P.R China
\\${}^e$ Center of Mathematical Science, Zhejiang
University, 38 Zheda Road Hangzhou, 310027 P.R China
\\
~~~~~~~\\
 \email{yjdu@fudan.edu.cn; b.feng@cms.zju.edu.cn; zhihaofu@nctu.edu.tw \hskip0.5cm} }
\preprint{LU-TP 14-11}

\date{\today}
\abstract{ We present an algorithm that leads to BCJ numerators satisfying  manifestly
the three properties proposed by Broedel and Carrasco in \cite{Broedel:2011pd}. We explicitly calculate the numerators at $4$,
$5$ and $6$-points and show that the relabeling property  is generically satisfied.
}


\begin{document}

\section{Introduction}

In an inspiring paper \cite{Bern:2008qj},  Bern, Carrasco
and Johansson (BCJ) have made a remarkable observation that the Yang-Mills scattering amplitude can be
rearranged into a symmetrical form where its kinematic dependent numerators
satisfy the same algebraic identities as the color factors. This duality
between color and kinematic factors was later found to be present
in a variety of Yang-Mills theories
\cite{Bern:2010ue,Carrasco:2011mn, Bern:2012uf, Boels:2012ew, Carrasco:2012ca, Boels:2013bi,  Bjerrum-Bohr:2013iza, Bern:2013yya, Chiodaroli:2013upa} and, perhaps most
surprisingly, was shown to be valid at least for the first few loop
levels \cite{Bern:2010yg, BjerrumBohr:2011xe, Du:2011js,Carrasco:2011mn, Boels:2011tp,Boels:2011mn, Du:2012mt, Oxburgh:2012zr, Saotome:2012vy, Boels:2012ew, Carrasco:2012ca, Boels:2013bi, Bjerrum-Bohr:2013iza, Bern:2013yya,  Nohle:2013bfa}. The apparently symmetrical structure also suggests
mirror versions of the existing formulations. In particular that studies
of the BCJ duals of the original color-ordered and the Del Duca-Dixon-Maltoni
formulations \cite{DelDuca:1999rs} can be found in \cite{Bern:2011rj, Bern:2011ia, Fu:2012uy,
 Du:2013sha,
Fu:2013qna, Du:2014uua}.

Ever since the discovery of the duality, a considerable amount of
endeavor has been devoted to the systematic construction of the kinematic
numerators. An explicit construction was given by Mafra, Schlotterer
and Stieberger using the pure spinor language\cite{Mafra:2011kj}. Alternatively,
it was shown that the kinematic factors can be interpreted in terms
of diffeomorphism algebra \cite{Monteiro:2011pc, BjerrumBohr:2012mg, Fu:2012uy, Monteiro:2013rya}. In a series of recent
papers \cite{Cachazo:2013gna, Cachazo:2013hca, Cachazo:2013iea} another interesting construction was provided by Cachazo, He
and Yuan (CHY) using the solutions to the scattering equations. At
the moment of writing it is not yet clear how to construct kinematic
numerators for the most generic configuration and to arbitrary loop
order.

Instead of attempting to decipher the analytic structure responsible
for the possible algebraic behavior, another line of thoughts is to
solve the kinematic numerators reversely in terms of scattering amplitudes,
and indeed, it was discussed in \cite{BjerrumBohr:2012mg, Boels:2012sy} that such expression
for the numerators can always be derived in suitable basis. A technical
issue lies with this approach is that because of the complexity involved,
along with the ambiguity introduced by generalized gauge invariance,
it is practically difficult to write down an analytic expression for
generic numerator. Nevertheless at tree level, explicit numerators
were worked out by Broedel and Carrasco \cite{Broedel:2011pd} at $4$ and
$5$-points, and $6$-points in the case of four dimensions, which
satisfy the following three properties:
\begin{enumerate}
\item The  numerators satisfy Jacobi identity, and this property is referred as {\it BCJ representation}.
\item All external state information such as particle species and helicity  is coded inside the color-ordered partial amplitudes. In other words, we 
have $n_i=\sum_{\a} c_{i\a} A_\a$, where $c_{i\a}$ is helicity blind. This property is called
{\it amplitude-encoded representation}.
\item  Expressions for numerators sharing the same topology  are
  relabeling related. In other words, for each topology, if we know
the BCJ numerator for a particular ordering of external legs, we know all others simply
through relabelings.
This property is called {\it symmetric representation}.
\end{enumerate}
In this paper we present a systematic construction of the kinematic
numerators at generic $n$-points via Kawai-Lewellen-Tye (KLT) relation \cite{KLT, Bern:1999bx}, which
satisfy the above three properties of Broedel and Carrasco.
As examples we derive the explicit formulas for $4$, $5$ and $6$ points.
In particular that the $6$-point expression here does not
assume spinor identities.

The paper is organized as follows.
In section \ref{sec:algorithm} we discuss the basic idea
used to determine the kinematic numerators. In sections \ref{sec:4pt}
and \ref{sec:5pt} we present explicit expressions for numerators
at $4$ and $5$-points. Due to   complexity we present the $6$-point result
in appendix \ref{sec:6pt-appd}. We show that the numerators
produced by the algorithm discussed in this paper satisfy all
three properties of  Broedel and Carrasco in section \ref{sec:relabeling}.
Finally, a conclusion is given in section \ref{sec:conclsn}.

\section{General framework}
\label{sec:algorithm}

Our starting point is the KLT   expression of the
total Yang-Mills amplitude \cite{KLT, Bern:1998ug, Bern:1999bx,  BjerrumBohr:2010ta},
\bea & & {\cal A}_n^{KLT}(1,2,...,n-1,n)= {\cal
A}_n^{KLT}(1,\{2,...,n-2\},n-1,n) \nn &  =  &
(-)^{n+1}\sum_{\sigma,\W\sigma\in S_{n-3}} A_n(1,\sigma(2,n-2),
n-1,n) {\cal S}[ \W\sigma(2,n-2))|\sigma(2,n-2))]_{p_1} \W
A_n(n-1,n,\W\sigma(2,n-2),1)\nn
& = &  (-)^{n+1}\sum_{\sigma,\W\sigma\in S_{n-3}}\W
A_n(1,n-1,n,\W\sigma(2,n-2)){\cal S}[
\W\sigma(2,n-2))|\sigma(2,n-2))]_{p_1}A_n(1,\sigma(2,n-2),
n-1,n)~~~\Label{KLT-Pure-S}\eea
where $A_n$ is the color ordered partial amplitude of Yang-Mills theory
and $\W A_n$ is the color ordered partial amplitude of scalar theory
appearing in the dual DDM-form. Notice that although the total amplitude 
is totally symmetric under permutations of all $n$ legs, the expressions in the second and
the third lines are not manifestly so. In particular that legs $1,n-1,n$ are kept fixed, while the
ordering of the rest   $(n-3)$ legs appear in ${\cal A}_n^{KLT}(1,2,...,n-1,n)$
has no effect on the expression. To emphasize this feature,
we write the parameter of amplitude ${\cal A}_n^{KLT}$ as $(1,\{2,...,n-2\},n-1,n)$.
The momentum kernel ${\cal S}$ appears in \eref{KLT-Pure-S} is defined as \cite{Bern:1998ug, BjerrumBohr:2010ta}
\bea {\cal S}[i_1,...,i_k|j_1,j_2,...,j_k]_{p_1} & = & \prod_{t=1}^k
(s_{i_t 1}+\sum_{q>t}^k \theta(i_t,i_q) s_{i_t
i_q})~~~\Label{S-def}\eea
where $\theta(i_t, i_q)$ is zero when  pair $(i_t,i_q)$ has same
ordering at both set ${\cal I}=\{i_1,...,i_k\},{\cal J}=\{j_1,j_2,...,j_k\}$, otherwise it is one.
A few examples are the 
following:
\bean {\cal S}[2,3,4|2,4,3] & = & s_{21} (s_{31}+s_{34}) s_{41},~~~
{\cal S}[2,3,4|4,3,2]=(s_{21}+s_{23}+s_{24})(s_{31}+s_{34}) s_{41}.
\eean
In this definition momentum $p_1$ plays a distinct role,
in the sense that for each leg $i$ there is always one term $s_{i 1}$. In the case when
different choices of $p_1$ are encountered, we should write
${\cal S}[{\cal I}|{\cal J}]_{p_1}$ to avoid  confusions.

Since our goal is to obtain a {\it symmetric representation} for BCJ numerators, as a second step
we symmetrize expression \eref{KLT-Pure-S}\footnote{Considering the average of KLT relations was firstly suggested in the work \cite{BjerrumBohr:2010hn}. However,
the numerator suggested in \cite{BjerrumBohr:2010hn} by taking $(n-2)!$ cannot satisfy the relabeling property.},

\bea {\cal A}_n^{S} & = & {1\over n!}\sum_{\sigma\in S_n} {\cal
A}_n^{KLT}(\sigma_1,\{\sigma_2,...,\sigma_{n-2}\},\sigma_{n-1},\sigma_n)
~~~\Label{M-total-n} \eea
However note that since KLT expression is already
manifestly $(n-3)!$-symmetric, equation \eref{M-total-n}
reduces to averaging over $n$-choices for $a_1$, then
$(n-1)$-choices for $a_{n-1}$ and $(n-2)$-choices for $a_n$,
\bea {\cal A}_n^{S} & = & {1\over n(n-1)(n-2)}{\sum_{i\neq j \neq k,
1}}^n {\cal A}_n^{KLT}(i,\{1,2,..,\WH i, ..., \WH j,.., \WH
k,...,n\},j,k)~~~\Label{M-total-n-1}\eea

As a third step, we expand $\W A_n$ using  KK-basis, with
, for example, $1$, $n$ fixed at the first and the last position.  Thus
${\cal A}_n^S$ becomes
\bea {\cal A}_n^S & = & \sum_{\sigma \in S_{n-2}} \W
n_{1\sigma(2,...,n-1)n} \W A_n (1,
\sigma(2,...,n-1),n),~~~~\Label{S-rep}\eea
where the $\W n_{1\sigma(2,...,n-1)n}$ here is a collective factor
of color-ordered amplitude $A_n$ and kinematic factors.
Comparing with the dual DDM-form \cite{Bern:2010yg} of the total Yang-Mills amplitude,
%
\bea {\cal A} = \sum_{\sigma\in S_{n-2}} n_{1\sigma n} \W
A(1,\sigma, n),~~~\Label{dual-DDM-form}\eea
 we propose that the desired
BCJ numerator $n_\a^S$ satisfying the three properties of \cite{Broedel:2011pd} to be
\bea n^s_{1\sigma(2,...,n-1)n}\equiv \W
n_{1\sigma(2,...,n-1)n}.~~~\Label{ns-def}\eea

To summarize, the basic idea is as follows:
\begin{itemize}
\item We construct only the KK-basis of BCJ numerator $n_i$ in \eref{dual-DDM-form}.
The rest numeators are given by  Jacobi identity.
By this construction, the expression is automatically  BCJ representation.

\item We consider constructing the numerators using KLT relation, where the helicity information is automatically coded inside the partial amplitude.

\item The ${\cal A}_n^S$ \eref{M-total-n} is averaged over all permutations of $n$ external legs, thus  the relabeling property is manifestly constructed.


\end{itemize}
%

\section{Four-point construction}
\label{sec:4pt}

Having discussed the general framework, let us present a few examples explicitly. We start with averaging
 KLT relation over all $4!$ permutations,
\bea
M(1,2,3,4)& =& {1\over 4!}\Sl_{\sigma\in S_{4}}M(\sigma)\nn
&=& \frac{-1}{24} \left(  A(1, 3, 4, 2)s_{12} \tilde{A}(1, 2, 3, 4) +
A(1, 4, 3, 2)s_{12} \tilde{A}(1, 2, 4, 3) \right. \nn
&&  \left. + \dots  + A(1, 3, 2, 4)s_{14} \tilde{A}(1, 4, 3, 2) \right)
.~~\Label{Average-tot-amp}
\eea
Translating all $\tilde{A}$ into KK-basis, and comparing equation \eref{Average-tot-amp}
with its dual DDM-form expression
\bea
M(1,2,3,4)=n_{1234}\W A(1,2,3,4)+n_{1324}\W A(1,3,2,4),
\eea
we see that numerator $n_{1234}$ can be read off from the coefficient of $\W A(1,2,3,4)$
in the symmetrized KLT expression  (\ref{Average-tot-amp}), giving
\bea n_{1234} &= & \frac{1}{24} \left( s_{12} A(2,3,4,1)-s_{12}
A(2,4,3,1)+s_{13}
   A(3,2,4,1)-s_{14} A(4,2,3,1)-s_{21} A(1,3,4,2) \right. \nn & & +s_{21}
   A(1,4,3,2)-s_{23} A(3,1,4,2)+s_{24} A(4,1,3,2)+s_{31}
   A(1,4,2,3)-s_{32} A(2,4,1,3)\nn & & +s_{34} A(4,1,2,3)-s_{34}
   A(4,2,1,3)-s_{41} A(1,3,2,4)+s_{42} A(2,3,1,4)-s_{43}
   A(3,1,2,4)\nn & & \left. +s_{43} A(3,2,1,4) \right) \Label{4pt-expression}\eea
The expression above can be further translated into BCJ basis,
\bea
n_{1234}&=&{1\over 3}[ s_{12}A(1,2,3,4)- s_{23}A(1,3,2,4)]={s_{12}(s_{13}-s_{23})\over 3s_{13}}A(1,2,3,4).~~\Label{4-pt-n-BCJbasis}
\eea
which is simply the four-point result obtained by  Broedel and Carrasco in \cite{Broedel:2011pd}.
The other numerator $n_{1324}$, obtained by collecting the coefficient of $\W A(1,3,2,4)$, gives
the same expression as (\ref{4pt-expression}) with legs $(2,3)$ swapped.


\section{The 5-point case}
\label{sec:5pt}
The calculation at $5$-points follows in a straightforward manner. Here we
list the result for $n_{12345}$ in KK-basis.
\begin{eqnarray}
n_{12345} & = &
\frac{1}{10} A(1, 2, 3, 4, 5)
   (s_{12}(s_{13}+s_{23}) +s_{45}(s_{34}+s_{35})) +
 \frac{1}{  10} A(1, 2, 4, 3, 5)
  ( -s_{12} s_{34} + s_{35} s_{45})
\nonumber \\ &&
+  \frac{1}{10 } A(1, 3, 2, 4, 5)
  ( s_{12} s_{13} - s_{45}s_{23} )
    \nonumber \\ &&
     +
\frac{1}{ 60} A(1, 3, 4, 2, 5)
  (s_{23} (s_{24} -  s_{25}) + s_{12} (3 s_{13} - s_{24} + s_{25}) +
 s_{13} (s_{14} - s_{15} + 2 s_{34} - 2 s_{35})
  \nonumber \\ && \hspace{3cm}
   +s_{45}(- 2 s_{14} -
  s_{45} + 3 s_{25}  + 2 s_{34}  + s_{35})\nn&&\hspace{3cm}
 -(s_{14} + s_{24} + s_{34}) (s_{14} + s_{23} + 2 (s_{24} + s_{34})))
 \nonumber \\ &&
  +
 \frac{1}{ 60} A(1, 4, 2, 3, 5)
  (s_{12} (s_{13} + 3 s_{14} - s_{15} + 2 s_{23} - 2 s_{25}) + s_{24} s_{34} -
 2 s_{13} s_{35} - s_{15} s_{35}
  \nonumber \\ && \hspace{3cm}
   + 2 s_{23} s_{35} + s_{25} s_{35} -
 s_{24} s_{45} + 3 s_{35} s_{45} + s_{14} (-s_{34} + s_{45}))
  \nonumber \\ &&
   +
\frac{1}{60} A(1, 4, 3, 2, 5)
  (s_{23} s_{24} + s_{12} (s_{13} + 2 s_{14} - s_{23} - 2 s_{25}) -
 s_{23} s_{25} - s_{14} s_{34}
 \nonumber \\ && \hspace{3cm}
 + 2 s_{23} s_{34} + s_{24} s_{34} +
 s_{13} (s_{14} - s_{15} - 2 s_{35}) - s_{15} s_{35} + s_{25} s_{35}
 \nonumber \\ && \hspace{3cm}
 -
 2 s_{14} s_{45} + 2 s_{25} s_{45} - s_{34} s_{45} + s_{35} s_{45}),
      \end{eqnarray}
which in BCJ basis simplifies to
\bea
n_{12345} & = &
\Biggl( A(1, 2, 3, 4, 5) (s_{1  4} + s_{2  4} +
      s_{3  4}) (s_{1
        3}^3 (2 s_{1  4} + 2 s_{2  3} - 3 (s_{2  4} + s_{3  4})) +  \Biggr.
         \nonumber \\ && \hspace{3cm}
      s_{1  3}^2 (3 s_{1  4}^2 + 3 s_{2  3}^2 - s_{2  4}^2 +
         3 s_{2  3} s_{3  4} - 6 s_{2  4} s_{3  4} - 5 s_{3  4}^2 -
         s_{1  4} (3 s_{2  4} + 2 s_{3  4}))
          \nonumber \\ && \hspace{3cm}
           -
      s_{1  4} s_{2
        3} (s_{1  4}^2 - s_{2  4}^2 + s_{2  4} s_{3  4} + 2 s_{3  4}^2 +
         3 s_{1  4} (s_{2  3} + s_{3  4}) +
         s_{2  3} (-2 s_{2  4} + 3 s_{3  4}))
          \nonumber \\ && \hspace{3cm}
           +
      s_{1  3} (s_{1  4}^3 - 3 s_{1  4}^2 s_{2  3} + s_{2  3}^3 +
         2 s_{2  3} s_{3  4} (s_{2  4} + s_{3  4}) +
         s_{2  3}^2 (2 s_{2  4} + 5 s_{3  4})
          \nonumber \\ && \hspace{3cm}
           -
         s_{3  4} (s_{2  4}^2 + 3 s_{2  4} s_{3  4} + 2 s_{3  4}^2) -
         s_{1  4} (3 s_{2  3}^2 + s_{2  4}^2 + 3 s_{2  4} s_{3  4} +
            3 s_{3  4}^2 + s_{2  3} (3 s_{2  4} + s_{3  4}))))
             \nonumber \\ &&
             +
   A(1, 2, 4, 3, 5) (s_{1  3}^4 s_{1  4} -
      s_{1  4} (s_{2  3} + s_{3  4}) (s_{1  4} + s_{2  4} +
         s_{3  4}) (s_{1  4}^2 - s_{2  4}^2 + s_{2  4} s_{3  4} +
         2 s_{3  4}^2
          \nonumber \\ && \hspace{3cm}
           + 3 s_{1  4} (s_{2  3} + s_{3  4}) +
         s_{2  3} (-2 s_{2  4} + 3 s_{3  4})) +
      s_{1
        3}^2 (2 s_{1  4}^3 +
         3 s_{1  4}^2 (s_{2  3} - s_{2  4} + 3 s_{3  4})
          \nonumber \\ && \hspace{3cm}
           -
         s_{2  4} (-3 s_{2  3}^2 + s_{2  4}^2 - 3 s_{2  3} s_{3  4} +
            6 s_{2  4} s_{3  4} + 5 s_{3  4}^2) +
             \nonumber \\ && \hspace{3cm}
         s_{1  4} (2 s_{2  3}^2 - 6 s_{2  4}^2 -
            3 s_{2  3} (s_{2  4} - 2 s_{3  4}) - 9 s_{2  4} s_{3  4} +
            7 s_{3  4}^2))
             \nonumber \\ && \hspace{3cm}
             +
      s_{1  3}^3 (3 s_{1  4}^2 +
         3 s_{1  4} (s_{2  3} - s_{2  4} + 2 s_{3  4}) +
         s_{2  4} (2 s_{2  3} - 3 (s_{2  4} + s_{3  4})))
          \nonumber \\ && \hspace{3cm}
          -
      s_{1  3} (s_{1  4}^3 (s_{2  3} - s_{3  4}) +
         s_{1  4}^2 (s_{2  3}^2 + (6 s_{2  4} - s_{3  4}) s_{3  4} +
            3 s_{2  3} (s_{2  4} + s_{3  4}))
             \nonumber \\ && \hspace{3cm}
              +
         s_{1  4} (s_{2  3}^2 (2 s_{2  4} + s_{3  4}) +
            s_{2  4} s_{3  4} (8 s_{2  4} + 9 s_{3  4}) +
            2 s_{2  3} (s_{2  4}^2 + 4 s_{2  4} s_{3  4} + s_{3  4}^2))
             \nonumber \\ && \hspace{3cm}
              \Biggl. +
         s_{2  4} (-s_{2  3}^3 - 2 s_{2  3} s_{3  4} (s_{2  4} + s_{3  4}) -
             s_{2  3}^2 (2 s_{2  4} + 5 s_{3  4}) +
            s_{3  4} (s_{2  4}^2 + 3 s_{2  4} s_{3  4} +
               2 s_{3  4}^2)))) \Biggr)
                \nonumber \\ &&
               \times \frac{1}{30 s_{1  3} s_{1
    4} (s_{1  3} + s_{1  4} + s_{3  4}))}
\eea
and we do find that
relabeling symmetry is satisfied at $5$-points. This result is also same
 with the one obtained by  Broedel and Carrasco in \cite{Broedel:2011pd}
The explicit expression for $6$-point numerator is
considerably more complicated and we leave the result to
appendix \ref{sec:6pt-appd}.

\section{Verifying symmetry properties of the numerators}
\label{sec:relabeling}

In this section let us check whether the BCJ numerators
constructed following the algorithm outlined at the beginning of this paper
indeed satisfy the three properties proposed in \cite{Broedel:2011pd}.
As remarked at the end of section \ref{sec:algorithm},
the $n_\a^s$'s defined by this algorithm satisfy the
BCJ-representation automatically since $n^s_{1\sigma(2,...,n-1)n}$
works as a basis and other numerators are constructed through antisymmetry and Jacobi
identity. The $n_\a^s$'s are also amplitude-encoded representation
since $n_\a^S$ is of the form $\sum A_n {\cal K}$, where ${\cal K}$ are
kinematic factors constructed by $s_{ij}$ and all helicity
information is included in $A_n$.

The last property, i.e., the symmetric representation, is however not
trivial.  Note that since all other topologies
can be constructed using  DDM-chains (linear trees), if we
can show the relabeling symmetry is true for this topology, it must be true for other
topologies. For the topology of DDM-chain, there are $n!$ different labelings.
Among them
 $(n-2)!$ of the numerators are directly given  by the algorithm \eref{ns-def}
 and others can be constructed
 using the Jacobi-identity and anti-symmetry.
In our third step, we have fixed only two legs $1,n$, the relabeling
property is manifestly true among $(2,3,...,n-1)$ by 
construction.
Since all permutations can be generated by
successive permutations between the $(n-1)$ consequtive pairs,
we can reduce our checking to
the following two permutations: $(12)$ and $(n-1)n$.
Now let us consider the DDM-chains $n_{213...(n-1)n}$ and $n_{123...n(n-1)}$.
We have two ways to
get the same $n_\a$ from basis numerators: One is by relabeling  $n_{123...(n-1)n}$
and another one, by using  Jacobi relation and antisymmetry.
If the expressions obtained by these two ways are the same, the relabeling property is  fully  checked.

\subsection{Permutation $(12)$}
Now we consider the relabeling
property under the permutation $(12)$ for $n_{123...(n-1)n}$. The
BCJ numerator $n_{213...(n-1)n}$ can be obtained by two ways. The
first way is by relabeling from $n_{123...(n-1)n}$,
which we denote as
%
\bea (n^S_{213...(n-1)n})_R \equiv
n^s_{123...(n-1)n}|_{1\leftrightarrow 2}~~~\Label{ns-1}\eea
The second way is by antisymmetry specially for this DDM-chain topology and we get
\bea (n^S_{213...(n-1)n})_A = -n^s_{123...(n-1)n}~~~\Label{ns-2}\eea
To check they are the same, we notice that permutation
$(12)$ in equation \eref{S-rep} gives 
\bea {\cal A}_n^S & = & \sum_{\sigma \in S_{n-2}} \W
n_{2\sigma(1,...,n-1)n} \W A_n (2,
\sigma(1,...,n-1),n)~~~~\Label{S-rep-1}\nn
& = & (n^S_{213...(n-1)n})_R\W A_n (2,1,3,...,n-1,n)+... \eea
 where we have written the expansion at the second line and identify
the coefficient of $\W A_n (2,1,3,...,n-1,n)$ to be the
$(n^S_{213...(n-1)n})_R$ given in \eref{ns-1}. The reason is simple:
because at our second step, symmetry among $n$ external legs are
manifest, thus when we go to different KK-basis, their coefficients
are related to each other by simple relabeling.

A relation between $(n^S_{213...(n-1)n})_R$ given in \eref{ns-1} and the original
$(1,n)$-basis numerators $n^S_{123...(n-1)n}$ can be obtained
by translating the amplitude $\W A(1,\sigma ,n)$ in equation \eref{S-rep}
into $(2,n)$-basis $\W A(2,\rho, n)$.
Notice however,
according to KK-relation,
%
\bea \W A(1,\a, 2,\b, n) & = &  \W A(2,\b, n, 1,\a)= (-)^{1+\#(\a)}
\sum_{ \rho \in COP(\b\bigcup (\a^T,1))} \W A(2, \rho,
n)~~~\Label{KK-exp-another} \eea
On the right hand side we see that
because of the presence of an extra $\a^T$ sitting on the left of leg $1$, we  never get
$\W A_n (2,1,3,...,n-1,n)$ unless  set $\a$ is empty.
When that happens,
there is only one contribution with a $(-)$ sign. Putting
\eref{KK-exp-another} back to \eref{S-rep} and we find the
coefficient of $\W A_n (2,1,3,...,n-1,n)$ is given by
$-n^s_{123...(n-1)n} $. Thus we have proven that \eref{ns-1} coincides with
\eref{ns-2}.

\subsection{Permutation $((n-1)n)$}
The proof of $((n-1)n)$ invariance is similar.
Using relabeling we find
%
\bea {\cal A}_n^S & = & \sum_{\sigma \in S_{n-2}} \W
n_{1\sigma(2,...,n-2,n)(n-1)} \W A_n (1,
\sigma(2,...,n-2,n),n-1)~~~~\Label{S-rep-2}\nn
& = & (n^S_{123...(n-2)n(n-1)})_R\W A_n (1,2,,...,n-2,n,n-1)+...
\eea
On the other hand expanding $\W A(1,\sigma, n)$ into $(1,n-1)$-basis gives
\bea \W A(1,\a, n-1,\b, n) & = & (-)^{1+\#(\b)} \sum_{ \rho \in
COP(\a\bigcup (n,\b^T,1))} \W A(1, \rho,
n-1)~~~\Label{KK-exp-another-1}\eea
we see on the right hand side of the equation that
because of the extra ordering $\b^T$ between legs $n$ and $1$, we  never get
$\W A_n (1,2,3,...,n-2,n,n-1)$ unless $\b$ is   empty. When this happens,
there is only one contribution with a $(-)$ sign, thus we
get
\bea (n^S_{123...(n-2)n(n-1)})_R=-n^s_{123...(n-1)n}  \eea
as required by anti-symmetry of the BCJ numerator.
\section{Conclusion}
\label{sec:conclsn}
In this paper we discussed a systematic construction of the BCJ numerator
based on matching KLT and dual DDM-form of the full Yang-Mills amplitude.
Using this method we explicitly calculated the numerators at $4$, $5$ and
$6$ points and verified the three symmetry properties proposed by Broedel
and Carrasco \cite{Broedel:2011pd} hold generically. Note the similarity between the
expressions discussed and the prescription proposed by Cachazo, He and Yuan
\cite{Cachazo:2013iea} despite the method used in this paper does not rely on the
existence of the solutions to scattering equations.

There are a few things one can proceed. First, although we have the general algorithm,
its computation takes a long time when the number of external leg increases. Thus it will be
nice if we can have a general patten of $n_\a$ expanded into the KK-basis (or BCJ-basis). Secondly,
it will be natural to generalize above results to loop level.
 At loop-level the
current method does not seem to straightforwardly generalize  because of the lack of support
from KLT relation, yet naively it might be possible to solve numerators
by comparing integrands in suitable basis. We leave this part of the
discussion to future works.

\section*{Acknowledgements}
Y. J. Du would like to acknowledge the EU programme Erasmus Mundus Action 2 and
the International Postdoctoral Exchange Fellowship Program of China(with Fudan University as the home university) for supporting his postdoctoral research in Lund University. Y. J. Du's research is supported in parts by the NSF of China Grant No.11105118, China Postdoctoral Science Foundation No.2013M530175 and the Fundamental Research Funds for the Central Universities of Fudan University No.20520133169.
C.F. would  like to acknowledge
the support from National Science Council, 50 billions project of
Ministry of Education and National Center for Theoretical Science,
Taiwan, Republic of China as well as the support from S.T. Yau
center of National Chiao Tung University.
B.F is supported, in part,
by fund from Qiu-Shi and Chinese NSF funding under contract
No.11031005, No.11135006, No. 11125523.

\appendix

\section{$6$-point numerator}
\label{sec:6pt-appd}
In this appendix we present the explicit formula
of the $6$-point BCJ numerator in KK basis.
Note however, despite its complixity, this expression
holds in all dimensions greater or equal to three
and does not depend on helicity except through color-ordered amplitudes.
\begin{eqnarray}
n_{123456} & = & \frac{1}{120} \left\{ A(123456) 2 \left[ 8 (s_{1  2} (s_{1  3} + s_{2  3}) (s_{1  4} + s_{2  4} +
      s_{3  4}) + (s_{3  4} + s_{3  5} + s_{3  6}) (s_{4  5} + s_{4  6}) s_{
     5  6}) \right] \right.
     \nonumber \\ &&
     +A(123546) 4 \left[ s_{1 2} (s_{1 3} + s_{2 3}) (s_{1 4} + s_{1 5} - s_{1 6}
          +   s_{2 4} + s_{2 5} - s_{2 6} + s_{3 4} + s_{3 5} - s_{3 6})
      \right.  \nonumber \\ &&\hspace{2cm} \left.
   + 2 (s_{3 4} + s_{3 5} + s_{3 6}) s_{4 6} s_{5 6} \right]
     \nonumber \\ &&
     +A(124356) 8 \left[  s_{3  5} + s_{3  6}) (s_{4  5} + s_{4  6}) s_{5  6} +
 4 s_{1  2} (2 s_{1  4} s_{2  3} + 2 s_{2  3} s_{2  4} +
    2 s_{1  3} (s_{1  4} + s_{2  4})
    \right.  \nonumber \\ &&\hspace{2cm} \left.
     - (s_{1  5} + s_{1  6} + s_{2  5} +
       s_{2  6}) s_{5  6}  \right]
     \nonumber \\ &&
     +A(124536) \left[ 8 s_{3  6}  (s_{4  5}  + s_{4  6} ) s_{5  6}  +
 s_{1  2}  (2 s_{1  5}  s_{2  4}  - 2 s_{1  6}  s_{2  4}  + 4 s_{2  3}  s_{2  4}  +
    2 s_{2  4}  s_{2  5}
    \right.  \nonumber \\ &&\hspace{2cm} \left.
     - 2 s_{2  4}  s_{2  6}  - s_{2  3}  s_{3  5}  +
    s_{2  3}  s_{3  6}  +
    s_{1  3}  (4 s_{1  4}  + 4 s_{2  4}  - s_{3  5}  + s_{3  6} ) +
    3 s_{2  4}  s_{4  5}
    \right.  \nonumber \\ &&\hspace{2cm} \left.
    +
    s_{1  4}  (2 s_{1  5}  - 2 s_{1  6}  + 4 s_{2  3}  + 2 s_{2  5}  -
       2 s_{2  6}  + 3 s_{4  5}  - 3 s_{4  6} ) - 3 s_{2  4}  s_{4  6}  -
    3 s_{1  5}  s_{5  6}
    \right.  \nonumber \\ &&\hspace{2cm} \left.
     - s_{1  6}  s_{5  6}  - 3 s_{2  5}  s_{5  6}  -
    s_{2  6}  s_{5  6} ) \right]
     \nonumber \\ &&
     +A(125346) \left[  8 (s_{3  4} + s_{3  6}) s_{4  6} s_{5  6} +
 s_{1  2} (4 s_{1  5} s_{2  3} - 2 s_{1  6} s_{2  3} + 2 s_{2  3} s_{2  4} +
    4 s_{2  3} s_{2  5}
    \right.  \nonumber \\ &&\hspace{2cm} \left.
    - 2 s_{2  3} s_{2  6} + 3 s_{2  3} s_{3  4} +
    s_{1  3} (2 s_{1  4} + 4 s_{1  5} - 2 s_{1  6} + 2 s_{2  4} +
       4 s_{2  5} - 2 s_{2  6} + 3 s_{3  4} - 3 s_{3  6})
       \right.  \nonumber \\ &&\hspace{2cm} \left.
        -
    3 s_{2  3} s_{3  6} - s_{1  5} s_{4  5} - s_{2  5} s_{4  5} +
    s_{1  4} (2 s_{2  3} - 3 s_{4  6}) - s_{1  6} s_{4  6} -
    3 s_{2  4} s_{4  6}
    \right.  \nonumber \\ &&\hspace{2cm} \left.
     - s_{2  6} s_{4  6} + s_{1  5} s_{5  6} +
    s_{2  5} s_{5  6}) \right]
     \nonumber \\ &&
     +A(125436) \left[ 8 s_{3  6} s_{4  6} s_{5  6} +
 s_{1  2} (2 s_{1  5} s_{2  3} + 2 s_{1  5} s_{2  4} - 2 s_{1  6} s_{2  4} +
    2 s_{2  3} s_{2  4} + 2 s_{2  3} s_{2  5} + 2 s_{2  4} s_{2  5}
    \right.  \nonumber \\ &&\hspace{2cm} \left.
     -
    2 s_{2  4} s_{2  6} - s_{2  3} s_{3  4} +
    s_{1  3} (2 s_{1  4} + 2 s_{1  5} + 2 s_{2  4} + 2 s_{2  5} -
       s_{3  4} - 3 s_{3  6}) - 3 s_{2  3} s_{3  6}
       \right.  \nonumber \\ &&\hspace{2cm} \left.
       - s_{1  5} s_{4  5} -
    s_{2  5} s_{4  5} +
    s_{1  4} (2 s_{1  5} - 2 s_{1  6} + 2 s_{2  3} + 2 s_{2  5} -
       2 s_{2  6}
              - 3 s_{4  6}) - s_{1  6} s_{4  6}
              \right.  \nonumber \\ &&\hspace{2cm} \left.
              - 3 s_{2  4} s_{4  6} -
    s_{2  6} s_{4  6} - 3 s_{1  5} s_{5  6} - 3 s_{2  5} s_{5  6}) \right]
     \nonumber \\ &&
     +A(132456) \left[ 4 (2 s_{1  2} s_{1
     3} (s_{1  4} + s_{2  4} + s_{3  4}) - (s_{1  4} + s_{1  5} + s_{1  6} -
       s_{2  4} - s_{2  5} - s_{2  6} - s_{3  4}
       \right.  \nonumber \\ &&\hspace{2cm} \left.
       - s_{3  5} -
      s_{3  6}) (s_{4  5} + s_{4  6}) s_{5  6}) \right]
     \nonumber \\ &&
     +A(132546) \left[ 4 (s_{1  2} s_{1
     3} (s_{1  4} + s_{1  5} - s_{1  6} + s_{2  4} + s_{2  5} - s_{2  6} +
      s_{3  4} + s_{3  5} - s_{3  6})
      \right.  \nonumber \\ &&\hspace{2cm} \left.
       + (-s_{1  4} - s_{1  5} - s_{1  6} +
      s_{2  4} + s_{2  5} + s_{2  6} + s_{3  4} + s_{3  5} + s_{3  6}) s_{4
     6} s_{5  6}) \right]
     \nonumber \\ &&
     +A(134256) \left[ -(s_{2  3} (s_{2  5} + s_{2  6}) +
     s_{1  3} (s_{1  5} + s_{1  6} +
        3 (s_{3  5} + s_{3  6})) + (3 s_{1  4} + 2 s_{1  5} + 2 s_{1  6}
        \right.  \nonumber \\ &&\hspace{2cm} \left.
         -
        4 s_{2  5} - 4 s_{2  6} - 3 s_{3  4} - 2 s_{3  5} -
        2 s_{3  6}) (s_{4  5} + s_{4  6})) s_{5  6} +
 s_{1  2} (8 s_{1  3} (s_{1  4} + s_{3  4})
 \right.  \nonumber \\ &&\hspace{2cm} \left.
  + (s_{2  5} + s_{2  6}) s_{5  6}) \right]
     \nonumber \\ &&
     +A(134526) \left[ -s_{2  3} s_{2  4} s_{2  5} + s_{2  3} s_{2  4} s_{2  6} -
 s_{2  3} s_{2  6} s_{5  6} - 2 s_{1  4} s_{4  5} s_{5  6} -
 s_{1  5} s_{4  5} s_{5  6} - s_{1  6} s_{4  5} s_{5  6}
 \right.  \nonumber \\ &&\hspace{2cm} \left.
 +
 4 s_{2  6} s_{4  5} s_{5  6} + 2 s_{3  4} s_{4  5} s_{5  6} +
 s_{3  5} s_{4  5} s_{5  6} + s_{3  6} s_{4  5} s_{5  6} -
 s_{1  4} s_{4  6} s_{5  6} - s_{1  5} s_{4  6} s_{5  6}
 \right.  \nonumber \\ &&\hspace{2cm} \left.
  -
 s_{1  6} s_{4  6} s_{5  6} + 4 s_{2  6} s_{4  6} s_{5  6} +
 s_{3  4} s_{4  6} s_{5  6} + s_{3  5} s_{4  6} s_{5  6} +
 s_{3  6} s_{4  6} s_{5  6} +
 s_{1  2} (s_{2  4} (s_{2  5} - s_{2  6})
 \right.  \nonumber \\ &&\hspace{2cm} \left.
 +
    s_{1  3} (4 s_{1  4} - s_{2  5} + s_{2  6} + 4 s_{3  4}) +
    s_{2  6} s_{5  6}) +
 s_{1  3} (-s_{1  6} s_{3  4} + s_{3  4} s_{3  5} - s_{3  4} s_{3  6}
 \right.  \nonumber \\ &&\hspace{2cm} \left.
  +
    2 s_{3  4} s_{4  5} +
    s_{1  4} (s_{1  5} - s_{1  6} + s_{3  5} - s_{3  6} + s_{4  5} -
       s_{4  6}) - 2 s_{3  4} s_{4  6} + s_{1  5} (s_{3  4} - s_{5  6})
       \right.  \nonumber \\ &&\hspace{2cm} \left.
        -
    2 s_{3  5} s_{5  6} - s_{3  6} s_{5  6})
    \right]
     \nonumber \\ &&
     +A(135246) \left[ s_{1  2} (s_{1
      3} (2 s_{1  4} + 4 s_{1  5} - 2 s_{1  6} + 3 s_{2  4} - 3 s_{2  6} +
       2 s_{3  4} + 4 s_{3  5} - 2 s_{3  6}) + s_{2  4} s_{4  6})
       \right.  \nonumber \\ &&\hspace{2cm} \left.
        -
 s_{1  3} ((s_{1  4} + 2 s_{3  4} + s_{3  6}) s_{4  6} +
    s_{3  5} (s_{4  5} - s_{5  6})) +
 s_{4  6} (-s_{2  3} s_{2
      4} + (-2 s_{1  4}
      \right.  \nonumber \\ &&\hspace{2cm} \left.
      - 3 s_{1  5} - 2 s_{1  6} + 4 s_{2  4} +
       4 s_{2  6} + 2 s_{3  4} + 3 s_{3  5} + 2 s_{3  6}) s_{5  6}) \right]
     \nonumber \\ &&
     +A(135426) \left[ -s_{2  3} s_{2  4} s_{2  5} + s_{2  3} s_{2  4} s_{2  6} -
 s_{2  3} s_{2  4} s_{4  5} - s_{2  3} s_{2  6} s_{5  6} +
 s_{1  5} s_{4  5} s_{5  6} - s_{3  5} s_{4  5} s_{5  6}
 \right.  \nonumber \\ &&\hspace{2cm} \left.
  -
 s_{1  4} s_{4  6} s_{5  6} - s_{1  5} s_{4  6} s_{5  6} -
 s_{1  6} s_{4  6} s_{5  6} + 4 s_{2  6} s_{4  6} s_{5  6} +
 s_{3  4} s_{4  6} s_{5  6} + s_{3  5} s_{4  6} s_{5  6}
 \right.  \nonumber \\ &&\hspace{2cm} \left.
  +
 s_{3  6} s_{4  6} s_{5  6} +
 s_{1  2} (s_{1
      3} (2 s_{1  4} + 2 s_{1  5} - s_{2  4} - 3 s_{2  6} + 2 s_{3  4} +
       2 s_{3  5})
       \right.  \nonumber \\ &&\hspace{2cm} \left.
       + s_{2  4} (s_{2  5} - s_{2  6} + s_{4  5}) +
    s_{2  6} s_{5  6}) +
 s_{1  3} (-s_{1  6} s_{3  4} + s_{3  4} s_{3  5} - s_{3  4} s_{3  6} -
    s_{3  5} s_{4  5}
    \right.  \nonumber \\ &&\hspace{2cm} \left.
    +
    s_{1  4} (s_{1  5} - s_{1  6} + s_{3  5} - s_{3  6} - s_{4  6}) -
    2 s_{3  4} s_{4  6} - s_{3  6} s_{4  6} +
    s_{1  5} (s_{3  4} - s_{5  6}) - 2 s_{3  5} s_{5  6}) \right]
      \nonumber \\ &&
     +A(142356) \left[ (-2 s_{1  6} s_{3  5} + 3 s_{2  3} s_{3  5} + 2 s_{2  5} s_{3  5} +
    2 s_{2  6} s_{3  5} - 2 s_{1  6} s_{3  6} + 3 s_{2  3} s_{3  6} +
    2 s_{2  5} s_{3  6} + 2 s_{2  6} s_{3  6}
    \right.  \nonumber \\ &&\hspace{2cm} \left.
     -
    3 s_{1  3} (s_{3  5} + s_{3  6}) - 2 s_{1  5} (s_{3  5} + s_{3  6}) +
    s_{1  4} s_{4  5} - s_{2  4} s_{4  5} + 4 s_{3  5} s_{4  5} +
    4 s_{3  6} s_{4  5} + s_{1  4} s_{4  6}
    \right.  \nonumber \\ &&\hspace{2cm} \left.
     - s_{2  4} s_{4  6} +
    4 s_{3  5} s_{4  6} + 4 s_{3  6} s_{4  6}) s_{5  6} +
 s_{1  2} (8 s_{1  3} s_{1  4} +
    8 s_{1  4} s_{2  3}
    \right.  \nonumber \\ &&\hspace{2cm} \left.
     - (s_{1  5} + s_{1  6} + 3 (s_{2  5} + s_{2  6})) s_{
      5  6}) \right]
     \nonumber \\ &&
     +A(142536) \left[ (-3 s_{1  3} s_{3  6} - 2 s_{1  5} s_{3  6} - 2 s_{1  6} s_{3  6} +
    3 s_{2  3} s_{3  6} + 2 s_{2  5} s_{3  6} + 2 s_{2  6} s_{3  6} +
    s_{1  4} s_{4  5} - s_{2  4} s_{4  5}
    \right.  \nonumber \\ &&\hspace{2cm} \left.
     + 4 s_{3  6} s_{4  5} +
    4 s_{3  6} s_{4  6}) s_{5  6} +
 s_{1  2} (4 s_{1  3} s_{1  4} - s_{2  3} s_{3  5} + s_{2  3} s_{3  6} +
    s_{1  4} (2 s_{1  5} - 2 s_{1  6}
    \right.  \nonumber \\ &&\hspace{2cm} \left.
    + 4 s_{2  3} + 2 s_{2  5} -
       2 s_{2  6} + 3 s_{4  5} - 3 s_{4  6}) - s_{1  5} s_{5  6} -
    2 s_{2  5} s_{5  6} - s_{2  6} s_{5  6}) \right]
\nonumber \\
&&
     +A(143256) \left[ -(s_{2  3} (s_{2  5} + s_{2  6}) + 2 s_{1  5} s_{3  5} +
     2 s_{1  6} s_{3  5} - 2 s_{2  5} s_{3  5} - 2 s_{2  6} s_{3  5} +
     2 s_{1  5} s_{3  6} + 2 s_{1  6} s_{3  6}
     \right.  \nonumber \\ &&\hspace{2cm} \left.
     - 2 s_{2  5} s_{3  6} -
     2 s_{2  6} s_{3  6} +
     s_{1  3} (s_{1  5} + s_{1  6} + 3 (s_{3  5} + s_{3  6})) +
     3 s_{1  4} s_{4  5} - 2 s_{2  5} s_{4  5}
     \right.  \nonumber \\ &&\hspace{2cm} \left.
      - 2 s_{2  6} s_{4  5} +
     s_{3  4} s_{4  5} - 2 s_{3  5} s_{4  5} - 2 s_{3  6} s_{4  5} +
     3 s_{1  4} s_{4  6} - 2 s_{2  5} s_{4  6} - 2 s_{2  6} s_{4  6} +
     s_{3  4} s_{4  6}
     \right.  \nonumber \\ &&\hspace{2cm} \left.
      - 2 s_{3  5} s_{4  6} - 2 s_{3  6} s_{4  6}) s_{5
   6} + s_{1  2} (8 s_{1  3} s_{1  4} - 3 (s_{2  5} + s_{2  6}) s_{5  6}) \right]
     \nonumber \\ &&
     +A(143526) \left[ -s_{2  3} s_{2  4} s_{2  5} + s_{2  3} s_{2  4} s_{2  6} -
 s_{2  3} s_{2  5} s_{3  4} + s_{2  3} s_{2  6} s_{3  4} -
 s_{1  4} s_{3  4} s_{4  5} + s_{1  4} s_{3  4} s_{4  6}
 \right.  \nonumber \\ &&\hspace{2cm} \left.
 -
 s_{2  3} s_{2  6} s_{5  6} - s_{1  5} s_{3  5} s_{5  6} -
 s_{1  6} s_{3  5} s_{5  6} + 2 s_{2  6} s_{3  5} s_{5  6} -
 s_{1  5} s_{3  6} s_{5  6} - s_{1  6} s_{3  6} s_{5  6}
 \right.  \nonumber \\ &&\hspace{2cm} \left.
 +
 2 s_{2  6} s_{3  6} s_{5  6} - 2 s_{1  4} s_{4  5} s_{5  6} +
 2 s_{2  6} s_{4  5} s_{5  6} - s_{3  4} s_{4  5} s_{5  6} +
 s_{3  5} s_{4  5} s_{5  6} + s_{3  6} s_{4  5} s_{5  6}
 \right.  \nonumber \\ &&\hspace{2cm} \left.
 -
 s_{1  4} s_{4  6} s_{5  6} + 2 s_{2  6} s_{4  6} s_{5  6} +
 s_{3  5} s_{4  6} s_{5  6} + s_{3  6} s_{4  6} s_{5  6} +
 s_{1  2} (4 s_{1  3} s_{1  4} + s_{2  3} s_{2  5}
 \right.  \nonumber \\ &&\hspace{2cm} \left.
 - s_{2  3} s_{2  6} +
    s_{1  4} (-s_{2  5} + s_{2  6}) - 3 s_{2  6} s_{5  6}) +
 s_{1  3} (s_{1
      4} (s_{1  5} - s_{1  6} + s_{3  5} - s_{3  6} + s_{4  5} -
       s_{4  6})
       \right.  \nonumber \\ &&\hspace{2cm} \left.
        - (s_{1  5} + 2 s_{3  5} + s_{3  6}) s_{5  6}) \right]
     \nonumber \\ &&
     +A(145236) \left[ -s_{2  3} s_{2  4} s_{3  6} - s_{2  3} s_{3  4} s_{3  6} -
 s_{1  4} s_{3  4} s_{4  5} - s_{1  4} s_{3  5} s_{4  5} -
 s_{1  4} s_{3  6} s_{4  6} - s_{1  5} s_{3  6} s_{5  6}
 \right.  \nonumber \\ &&\hspace{2cm} \left.
  -
 s_{1  6} s_{3  6} s_{5  6} + 3 s_{2  3} s_{3  6} s_{5  6} +
 2 s_{2  6} s_{3  6} s_{5  6} + 2 s_{1  4} s_{4  5} s_{5  6} -
 s_{2  4} s_{4  5} s_{5  6} - s_{2  5} s_{4  5} s_{5  6}
 \right.  \nonumber \\ &&\hspace{2cm} \left.
  +
 3 s_{3  6} s_{4  5} s_{5  6} + 2 s_{3  6} s_{4  6} s_{5  6} -
 s_{1  3} s_{3  6} (s_{1  4} + s_{5  6}) +
 s_{1  2} (2 s_{1  3} s_{1  4} - s_{2  3} s_{2  5} - s_{2  3} s_{3  5}
 \right.  \nonumber \\ &&\hspace{2cm} \left.
  +
    2 s_{2  3} s_{3  6} +
    s_{1  4} (2 s_{1  5} - s_{1  6} + 3 s_{2  3} - s_{2  6} + 3 s_{4  5} -
       s_{4  6}) - s_{1  5} s_{5  6} - s_{2  6} s_{5  6}) \right]
     \nonumber \\ &&
     +A(145326) \left[ -s_{2  3} s_{2  4} s_{2  5} + s_{2  3} s_{2  4} s_{2  6} -
 s_{2  3} s_{2  5} s_{3  4} + s_{2  3} s_{2  6} s_{3  4} -
 s_{2  3} s_{2  4} s_{3  5} - s_{2  3} s_{3  4} s_{3  5}
 \right.  \nonumber \\ &&\hspace{2cm} \left.
  -
 s_{1  4} s_{3  4} s_{4  5} - s_{1  4} s_{3  5} s_{4  5} -
 s_{1  4} s_{3  6} s_{4  6} - s_{2  3} s_{2  6} s_{5  6} -
 s_{1  5} s_{3  6} s_{5  6} - s_{1  6} s_{3  6} s_{5  6}
 \right.  \nonumber \\ &&\hspace{2cm} \left.
  +
 2 s_{2  6} s_{3  6} s_{5  6} - 2 s_{1  4} s_{4  5} s_{5  6} +
 2 s_{2  6} s_{4  5} s_{5  6} - s_{3  4} s_{4  5} s_{5  6} -
 s_{3  5} s_{4  5} s_{5  6} + s_{3  6} s_{4  5} s_{5  6}
 \right.  \nonumber \\ &&\hspace{2cm} \left.
 +
 s_{2  6} s_{4  6} s_{5  6} + s_{3  6} s_{4  6} s_{5  6} +
 s_{1  2} (2 s_{1  3} s_{1  4} + s_{2  3} s_{2  5} - 2 s_{2  3} s_{2  6} +
    s_{2  3} s_{3  5}
    \right.  \nonumber \\ &&\hspace{2cm} \left.
     +
    s_{1  4} (s_{1  5} - s_{2  3} - s_{2  6} + 2 s_{4  5}) -
    s_{2  6} s_{5  6}) +
 s_{1  3} (s_{1
      4} (s_{1  5} - s_{1  6} - s_{3  6} + s_{4  5} -
       s_{4  6})
       \right.  \nonumber \\ &&\hspace{2cm} \left.
        - (s_{1  5} + s_{3  6}) s_{5  6}) \right]
      \nonumber \\ &&
     +A(152346) \left[ -s_{2  5} s_{3  5} s_{4  5} - 2 s_{1  3} s_{3  4} s_{4  6} -
 s_{1  4} s_{3  4} s_{4  6} - s_{1  6} s_{3  4} s_{4  6} +
 2 s_{2  3} s_{3  4} s_{4  6} + s_{2  4} s_{3  4} s_{4  6}
 \right.  \nonumber \\ &&\hspace{2cm} \left.
  +
 s_{2  6} s_{3  4} s_{4  6} - s_{1  3} s_{3  6} s_{4  6} -
 s_{1  4} s_{3  6} s_{4  6} - s_{1  6} s_{3  6} s_{4  6} +
 s_{2  3} s_{3  6} s_{4  6} + s_{2  4} s_{3  6} s_{4  6}
 \right.  \nonumber \\ &&\hspace{2cm} \left.
 +
 s_{2  6} s_{3  6} s_{4  6} + s_{2  5} s_{3  5} s_{5  6} -
 s_{2  5} s_{4  6} s_{5  6} + 4 s_{3  4} s_{4  6} s_{5  6} +
 4 s_{3  6} s_{4  6} s_{5  6}
 \right.  \nonumber \\ &&\hspace{2cm} \left.
 +
 s_{1  2} (4 s_{1  5} s_{2  3} - s_{1  6} s_{2  3} + s_{2  3} s_{2  4} -
    s_{2  3} s_{2  6} + 2 s_{2  3} s_{3  4} +
    s_{1  3} (s_{1  4} + 4 s_{1  5} - s_{1  6}
    \right.  \nonumber \\ &&\hspace{2cm} \left.
     + s_{2  4} - s_{2  6} +
       s_{3  4} - s_{3  6}) - 2 s_{2  3} s_{3  6} - s_{1  5} s_{4  5} +
    s_{1  4} (s_{2  3} - s_{4  6}) - 2 s_{2  4} s_{4  6}
    \right.  \nonumber \\ &&\hspace{2cm} \left.
     -
    s_{2  6} s_{4  6} + s_{1  5} s_{5  6}) +
 s_{1  5} (s_{3  5} (s_{4  5} - s_{5  6}) + s_{4  6} s_{5  6})\right]
     \nonumber \\ &&
     +A(152436) \left[ -s_{2  3} s_{3  4} s_{3  6} + s_{1  5} s_{3  4} s_{4  5} -
 s_{2  5} s_{3  4} s_{4  5} + s_{1  5} s_{3  5} s_{4  5} -
 s_{2  5} s_{3  5} s_{4  5} + s_{1  3} s_{3  6} (s_{3  4} - s_{4  6})
 \right.  \nonumber \\ &&\hspace{2cm} \left.
  -
 s_{1  4} s_{3  6} s_{4  6} - s_{1  6} s_{3  6} s_{4  6} +
 s_{2  3} s_{3  6} s_{4  6} + s_{2  4} s_{3  6} s_{4  6} +
 s_{2  6} s_{3  6} s_{4  6} + s_{1  5} s_{3  6} s_{5  6}
 \right.  \nonumber \\ &&\hspace{2cm} \left.
 -
 s_{2  5} s_{3  6} s_{5  6} - s_{1  5} s_{4  5} s_{5  6} +
 s_{2  5} s_{4  5} s_{5  6} + 4 s_{3  6} s_{4  6} s_{5  6} +
 s_{1  2} (2 s_{1  5} s_{2  3} + 2 s_{1  5} s_{2  4}
 \right.  \nonumber \\ &&\hspace{2cm} \left.
 - s_{1  6} s_{2  4} +
    s_{2  3} s_{2  4} - s_{2  4} s_{2  6} - s_{2  3} s_{3  4} +
    s_{1  3} (s_{1  4} + 2 s_{1  5} + s_{2  4} - s_{3  6}) -
    2 s_{2  3} s_{3  6}
    \right.  \nonumber \\ &&\hspace{2cm} \left.
    - s_{1  5} s_{4  5} +
    s_{1  4} (2 s_{1  5} - s_{1  6} + s_{2  3} - s_{2  6} - s_{4  6}) -
    2 s_{2  4} s_{4  6} - s_{2  6} s_{4  6} - 3 s_{1  5} s_{5  6}) \right]
     \nonumber \\ &&
     +A(153246) \left[ -s_{2  3} s_{3  5} s_{4  5} - s_{2  5} s_{3  5} s_{4  5} -
 s_{1  3} s_{1  4} s_{4  6} - s_{2  3} s_{2  4} s_{4  6} -
 2 s_{1  3} s_{3  4} s_{4  6} - s_{1  4} s_{3  4} s_{4  6}
 \right.  \nonumber \\ &&\hspace{2cm} \left.
  -
 s_{1  6} s_{3  4} s_{4  6} + s_{2  4} s_{3  4} s_{4  6} +
 s_{2  6} s_{3  4} s_{4  6} - s_{1  3} s_{3  6} s_{4  6} -
 s_{1  4} s_{3  6} s_{4  6} - s_{1  6} s_{3  6} s_{4  6}
 \right.  \nonumber \\ &&\hspace{2cm} \left.
  +
 s_{2  4} s_{3  6} s_{4  6} + s_{2  6} s_{3  6} s_{4  6} +
 s_{2  3} s_{3  5} s_{5  6} + s_{2  5} s_{3  5} s_{5  6} +
 2 s_{2  4} s_{4  6} s_{5  6} + 2 s_{2  6} s_{4  6} s_{5  6}
 \right.  \nonumber \\ &&\hspace{2cm} \left.
 +
 2 s_{3  4} s_{4  6} s_{5  6} - s_{3  5} s_{4  6} s_{5  6} +
 2 s_{3  6} s_{4  6} s_{5  6} +
 s_{1  2} (s_{2  3} (-s_{2  4} + s_{2  6})
 \right.  \nonumber \\ &&\hspace{2cm} \left.
  +
    s_{1  3} (s_{1  4} + 4 s_{1  5} - s_{1  6} + s_{2  4} - s_{2  6} +
       s_{3  4} - s_{3  6}) - s_{1  5} s_{4  5} - 2 s_{2  4} s_{4  6} -
    s_{2  6} s_{4  6} + s_{1  5} s_{5  6})
    \right.  \nonumber \\ &&\hspace{2cm} \left.
    +
 s_{1  5} (s_{3  5} (s_{4  5} - s_{5  6}) - 3 s_{4  6} s_{5  6}) \right]
     \nonumber \\ &&
     +A(153426) \left[ -s_{2  3} s_{2  4} s_{2  5} + s_{2  3} s_{2  4} s_{2  6} -
 s_{2  3} s_{2  4} s_{3  5} - s_{2  3} s_{2  4} s_{4  5} +
 s_{1  5} s_{3  5} s_{4  5} - s_{2  3} s_{3  5} s_{4  5}
 \right.  \nonumber \\ &&\hspace{2cm} \left.
  -
 s_{2  4} s_{3  5} s_{4  5} - s_{2  5} s_{3  5} s_{4  5} -
 s_{1  4} s_{3  4} s_{4  6} - s_{1  6} s_{3  4} s_{4  6} +
 s_{2  6} s_{3  4} s_{4  6} - s_{1  4} s_{3  6} s_{4  6}
 \right.  \nonumber \\ &&\hspace{2cm} \left.
  -
 s_{1  6} s_{3  6} s_{4  6} + s_{2  6} s_{3  6} s_{4  6} +
 s_{1  2} (s_{2  3} s_{2  4} + s_{2  3} s_{2  6} + s_{2  4} s_{2  6} +
    s_{1  3} (s_{1  4} + 2 s_{1  5}
    \right.  \nonumber \\ &&\hspace{2cm} \left.
     - s_{2  6} + s_{3  4}) -
    s_{1  5} (s_{2  4} + s_{4  5}) - s_{2  6} s_{4  6}) +
 s_{1  3} (-s_{1  6} s_{3  4} - s_{3  4} s_{3  6}
 \right.  \nonumber \\ &&\hspace{2cm} \left.
  +
    s_{1  4} (s_{1  5} - s_{1  6} - s_{3  6} - s_{4  6}) -
    2 s_{3  4} s_{4  6} - s_{3  6} s_{4  6} +
    s_{1  5} (s_{3  4} - s_{5  6}))
    \right.  \nonumber \\ &&\hspace{2cm} \left.
     - s_{2  3} s_{2  6} s_{5  6} +
 s_{1  5} s_{3  5} s_{5  6} - s_{2  6} s_{3  5} s_{5  6} +
 s_{1  5} s_{4  5} s_{5  6} + s_{3  5} s_{4  5} s_{5  6} -
 s_{1  5} s_{4  6} s_{5  6}
 \right.  \nonumber \\ &&\hspace{2cm} \left.
 + 2 s_{2  6} s_{4  6} s_{5  6} +
 s_{3  4} s_{4  6} s_{5  6} + s_{3  6} s_{4  6} s_{5  6} \right]
\nonumber \\
 &&
     +A(154236) \left[  -s_{2  3} s_{2  4} s_{3  6} - s_{2  3} s_{3  4} s_{3  6} +
 s_{1  5} s_{3  4} s_{4  5} - s_{2  4} s_{3  4} s_{4  5} -
 s_{2  5} s_{3  4} s_{4  5} + s_{1  5} s_{3  5} s_{4  5}
 \right.  \nonumber \\ &&\hspace{2cm} \left.
  -
 s_{2  4} s_{3  5} s_{4  5} - s_{2  5} s_{3  5} s_{4  5} -
 s_{1  4} s_{3  6} s_{4  6} - s_{1  6} s_{3  6} s_{4  6} +
 s_{2  3} s_{3  6} s_{4  6} + s_{2  6} s_{3  6} s_{4  6}
 \right.  \nonumber \\ &&\hspace{2cm} \left.
 -
 s_{1  3} s_{3  6} (s_{1  4} + s_{4  6}) - s_{1  5} s_{3  6} s_{5  6} +
 2 s_{2  3} s_{3  6} s_{5  6} + s_{2  6} s_{3  6} s_{5  6} -
 2 s_{1  5} s_{4  5} s_{5  6}
 \right.  \nonumber \\ &&\hspace{2cm} \left.
  + s_{2  4} s_{4  5} s_{5  6} +
 s_{2  5} s_{4  5} s_{5  6} - s_{3  6} s_{4  5} s_{5  6} +
 2 s_{3  6} s_{4  6} s_{5  6} +
 s_{1  2} (s_{1  3} (s_{1  4} + s_{1  5})
 \right.  \nonumber \\ &&\hspace{2cm} \left.
 + 2 s_{1  5} s_{2  3} -
    s_{2  3} s_{2  4} - s_{2  3} s_{3  4} - 2 s_{2  3} s_{3  6} -
    s_{1  5} s_{4  5} +
    s_{1  4} (2 s_{1  5} - s_{1  6} + s_{2  3} - s_{2  6} - s_{4  6})
    \right.  \nonumber \\ &&\hspace{2cm} \left.
    -
    s_{2  6} s_{4  6} - s_{1  5} s_{5  6}) \right]
     \nonumber \\ &&
     +A(154326)     \left[
     -s_{2  3} s_{2  4} s_{2  5} + s_{2  3} s_{2  4} s_{2  6} -
 s_{2  3} s_{2  5} s_{3  4} + s_{2  3} s_{2  6} s_{3  4} -
 s_{2  3} s_{2  4} s_{3  5} - s_{2  3} s_{3  4} s_{3  5}
 \right.  \nonumber \\ &&\hspace{2cm} \left.
 -
 s_{2  3} s_{2  4} s_{4  5} + s_{1  5} s_{3  4} s_{4  5} -
 2 s_{2  3} s_{3  4} s_{4  5} - s_{2  4} s_{3  4} s_{4  5} -
 s_{2  5} s_{3  4} s_{4  5} + s_{1  5} s_{3  5} s_{4  5}
 \right.  \nonumber \\ &&\hspace{2cm} \left.
  -
 s_{2  3} s_{3  5} s_{4  5} - s_{2  4} s_{3  5} s_{4  5} -
 s_{2  5} s_{3  5} s_{4  5} - s_{1  4} s_{3  6} s_{4  6} -
 s_{1  6} s_{3  6} s_{4  6} + s_{2  6} s_{3  6} s_{4  6}
 \right.  \nonumber \\ &&\hspace{2cm} \left.
  +
 s_{1  2} (s_{1  3} (s_{1  4} + s_{1  5}) - s_{1  5} s_{2  3} +
    s_{2  3} s_{2  4} + s_{1  4} (s_{1  5} - s_{2  6}) +
    2 s_{2  3} s_{2  6} + s_{2  3} s_{3  4}
    \right.  \nonumber \\ &&\hspace{2cm} \left.
     - s_{1  5} s_{4  5} -
    s_{2  6} s_{4  6}) - s_{2  3} s_{2  6} s_{5  6} -
 s_{1  5} s_{3  6} s_{5  6} + s_{2  6} s_{3  6} s_{5  6} +
 2 s_{1  5} s_{4  5} s_{5  6}
 \right.  \nonumber \\ &&\hspace{2cm} \left.
  - s_{2  6} s_{4  5} s_{5  6} +
 s_{3  4} s_{4  5} s_{5  6} + s_{3  5} s_{4  5} s_{5  6} +
 s_{2  6} s_{4  6} s_{5  6} + s_{3  6} s_{4  6} s_{5  6}
 \right.  \nonumber \\ &&\hspace{2cm} \left.
 \left.
  +
 s_{1  3} (s_{1  4} (s_{1  5} - s_{1  6} - s_{3  6} - s_{4  6}) -
    s_{3  6} s_{4  6} - s_{1  5} s_{5  6})
    \right]    \right\}
     \end{eqnarray}


\begin{thebibliography}{References}




\bibitem{Bern:2008qj}
  Z.~Bern, J.~J.~M.~Carrasco and H.~Johansson,
  ``New Relations for Gauge-Theory Amplitudes,''
  Phys.\ Rev.\ D {\bf 78}  (2008) 085011
  [arXiv:0805.3993 [hep-ph]].

\bibitem{Bern:2010ue}
  Z.~Bern, J.~J.~M.~Carrasco and H.~Johansson,
  ``Perturbative Quantum Gravity as a Double Copy of Gauge Theory,''
  Phys.\ Rev.\ Lett.\  {\bf 105}  (2010) 061602
  [arXiv:1004.0476 [hep-th]].

\bibitem{Carrasco:2011mn}
  J.~J.~.Carrasco and H.~Johansson,
  ``Five-Point Amplitudes in N=4 Super-Yang-Mills Theory and N=8 Supergravity,''
  Phys.\ Rev.\ D {\bf 85} (2012) 025006
  [arXiv:1106.4711 [hep-th]].



\bibitem{Bern:2012uf}
  Z.~Bern, J.~J.~M.~Carrasco, L.~J.~Dixon, H.~Johansson and R.~Roiban,
  ``Simplifying Multiloop Integrands and Ultraviolet Divergences of Gauge Theory and Gravity Amplitudes,''
  Phys.\ Rev.\ D {\bf 85} (2012) 105014
  [arXiv:1201.5366 [hep-th]].


\bibitem{Boels:2012ew}
  R.~H.~Boels, B.~A.~Kniehl, O.~V.~Tarasov and G.~Yang,
  ``Color-kinematic Duality for Form Factors,''
  JHEP {\bf 1302} (2013) 063
  [arXiv:1211.7028 [hep-th]].

\bibitem{Carrasco:2012ca}
  J.~J.~M.~Carrasco, M.~Chiodaroli, M.~G¨¹naydin and R.~Roiban,
  ``One-loop four-point amplitudes in pure and matter-coupled N $\leq$ 4 supergravity,''
  JHEP {\bf 1303} (2013) 056
  [arXiv:1212.1146 [hep-th]].



\bibitem{Boels:2013bi}
  R.~H.~Boels, R.~S.~Isermann, R.~Monteiro and D.~O'Connell,
  ``Colour-Kinematics Duality for One-Loop Rational Amplitudes,''
  JHEP {\bf 1304} (2013) 107
  [arXiv:1301.4165 [hep-th]].

\bibitem{Bjerrum-Bohr:2013iza}
  N.~E.~J.~Bjerrum-Bohr, T.~Dennen, R.~Monteiro and D.~O'Connell,
  ``Integrand Oxidation and One-Loop Colour-Dual Numerators in N=4 Gauge Theory,''
  JHEP {\bf 1307} (2013) 092
  [arXiv:1303.2913 [hep-th]].

\bibitem{Bern:2013yya}
  Z.~Bern, S.~Davies, T.~Dennen, Y.~-t.~Huang and J.~Nohle,
  ``Color-Kinematics Duality for Pure Yang-Mills and Gravity at One and Two Loops,''
  arXiv:1303.6605 [hep-th].



















\bibitem{Chiodaroli:2013upa}
  M.~Chiodaroli, Q.~Jin and R.~Roiban,
  ``Color/kinematics duality for general abelian orbifolds of N=4 super Yang-Mills theory,''
  JHEP {\bf 1401} (2014) 152
  [arXiv:1311.3600 [hep-th], arXiv:1311.3600].

\bibitem{Bern:2010yg}
  Z.~Bern, T.~Dennen, Y.~-t.~Huang and M.~Kiermaier,
  ``Gravity as the Square of Gauge Theory,''
  Phys.\ Rev.\ D {\bf 82}  (2010) 065003
  [arXiv:1004.0693 [hep-th]].

\bibitem{BjerrumBohr:2011xe}
  N.~E.~J.~Bjerrum-Bohr, P.~H.~Damgaard, H.~Johansson and T.~Sondergaard,
  ``Monodromy--like Relations for Finite Loop Amplitudes,''
  JHEP {\bf 1105}, 039 (2011)
  [arXiv:1103.6190 [hep-th]].



\bibitem{Du:2011js}
  Y.~-J.~Du, B.~Feng and C.~-H.~Fu,
  ``BCJ Relation of Color Scalar Theory and KLT Relation of Gauge Theory,''
  JHEP {\bf 1108} (2011) 129
  [arXiv:1105.3503 [hep-th]].





















\bibitem{Boels:2011tp}
  R.~H.~Boels and R.~S.~Isermann,
  ``New relations for scattering amplitudes in Yang-Mills theory at loop level,''
  Phys.\ Rev.\ D {\bf 85} (2012) 021701
  [arXiv:1109.5888 [hep-th]].

\bibitem{Boels:2011mn}
  R.~H.~Boels and R.~S.~Isermann,
  ``Yang-Mills amplitude relations at loop level from non-adjacent BCFW shifts,''
  JHEP {\bf 1203} (2012) 051
  [arXiv:1110.4462 [hep-th]].


\bibitem{Du:2012mt}
  Y.~-J.~Du and H.~Luo,
  ``On General BCJ Relation at One-loop Level in Yang-Mills Theory,''
  JHEP {\bf 1301} (2013) 129
  [arXiv:1207.4549 [hep-th]].




\bibitem{Oxburgh:2012zr}
  S.~Oxburgh and C.~D.~White,
  ``BCJ duality and the double copy in the soft limit,''
  JHEP {\bf 1302} (2013) 127
  [arXiv:1210.1110 [hep-th]].

\bibitem{Saotome:2012vy}
  R.~Saotome and R.~Akhoury,
  ``Relationship Between Gravity and Gauge Scattering in the High Energy Limit,''
  JHEP {\bf 1301} (2013) 123
   [JHEP {\bf 1301} (2013) 123]
  [arXiv:1210.8111 [hep-th]].






\bibitem{Nohle:2013bfa}
  J.~Nohle,
  ``Color-Kinematics Duality in One-Loop Four-Gluon Amplitudes with Matter,''
  arXiv:1309.7416 [hep-th].




\bibitem{DelDuca:1999rs}
  V.~Del Duca, L.~J.~Dixon and F.~Maltoni,
  ``New color decompositions for gauge amplitudes at tree and loop level,''
  Nucl.\ Phys.\ B {\bf 571} (2000) 51
  [hep-ph/9910563].


\bibitem{Bern:2011rj}
  Z.~Bern, C.~Boucher-Veronneau and H.~Johansson,
  ``N $\geq$ 4 Supergravity Amplitudes from Gauge Theory at One Loop,''
  Phys.\ Rev.\ D {\bf 84} (2011) 105035
  [arXiv:1107.1935 [hep-th]].


\bibitem{Bern:2011ia}
  Z.~Bern and T.~Dennen,
  ``A Color Dual Form for Gauge-Theory Amplitudes,''  Phys.\ Rev.\ Lett.\
   {\bf 107}, 081601 (2011)  [arXiv:1103.0312 [hep-th]].  


\bibitem{Fu:2012uy}
  C.~-H.~Fu, Y.~-J.~Du and B.~Feng,
  ``An algebraic approach to BCJ numerators,''
  JHEP {\bf 1303} (2013) 050
  [arXiv:1212.6168 [hep-th]].





\bibitem{Du:2013sha}
  Y.~-J.~Du, B.~Feng and C.~-H.~Fu,
  ``The Construction of Dual-trace Factor in Yang-Mills Theory,''
  JHEP {\bf 1307} (2013) 057
  [arXiv:1304.2978 [hep-th]].


\bibitem{Fu:2013qna}
  C.~-H.~Fu, Y.~-J.~Du and B.~Feng,
  ``Note on Construction of Dual-trace Factor in Yang-Mills Theory,''
  JHEP {\bf 1310} (2013) 069
  [arXiv:1305.2996 [hep-th]].


\bibitem{Du:2014uua}
  Y.~-J.~Du, B.~Feng and C.~-H.~Fu,
  ``Dual-color decompositions at one-loop level in Yang-Mills theory,''
  arXiv:1402.6805 [hep-th].



\bibitem{Mafra:2011kj}
  C.~R.~Mafra, O.~Schlotterer and S.~Stieberger,
  ``Explicit BCJ Numerators from Pure Spinors,''
  JHEP {\bf 1107} (2011) 092
  [arXiv:1104.5224 [hep-th]].



\bibitem{Monteiro:2011pc}
  R.~Monteiro and D.~O'Connell,
  ``The Kinematic Algebra From the Self-Dual Sector,''
  JHEP {\bf 1107} (2011) 007
  [arXiv:1105.2565 [hep-th]].

\bibitem{BjerrumBohr:2012mg}
  N.~E.~J.~Bjerrum-Bohr, P.~H.~Damgaard, R.~Monteiro and D.~O'Connell,
  ``Algebras for Amplitudes,''
  JHEP {\bf 1206} (2012) 061
  [arXiv:1203.0944 [hep-th]].

\bibitem{Monteiro:2013rya}
  R.~Monteiro and D.~O'Connell,
  ``The Kinematic Algebras from the Scattering Equations,''
  arXiv:1311.1151 [hep-th].

\bibitem{Cachazo:2013gna}
  F.~Cachazo, S.~He and E.~Y.~Yuan,
  ``Scattering Equations and KLT Orthogonality,''
  arXiv:1306.6575 [hep-th].

\bibitem{Cachazo:2013hca}
  F.~Cachazo, S.~He and E.~Y.~Yuan,
  ``Scattering of Massless Particles in Arbitrary Dimension,''
  arXiv:1307.2199 [hep-th].


\bibitem{Cachazo:2013iea}
  F.~Cachazo, S.~He and E.~Y.~Yuan,
 ``Scattering of Massless Particles: Scalars, Gluons and Gravitons,''
  arXiv:1309.0885 [hep-th].



\bibitem{Boels:2012sy}
  R.~H.~Boels and R.~S.~Isermann,
  ``On powercounting in perturbative quantum gravity theories through color-kinematic duality,''
  JHEP {\bf 1306} (2013) 017
  [arXiv:1212.3473].



\bibitem{Broedel:2011pd}
  J.~Broedel and J.~J.~M.~Carrasco,
  ``Virtuous Trees at Five and Six Points for Yang-Mills and Gravity,''
  Phys.\ Rev.\ D {\bf 84} (2011) 085009
  [arXiv:1107.4802 [hep-th]].


  \bibitem{KLT} H. Kawai, D. Lewellen and H. Tye, "A Relation Betwwen Tree
Amplitudes of Closed and Open Strings", Nucl.Phys.B269 (1986)1.

\bibitem{Bern:1999bx}
  Z.~Bern, A.~De Freitas and H.~L.~Wong,
  ``On the coupling of gravitons to matter,''
  Phys.\ Rev.\ Lett.\  {\bf 84} (2000) 3531
  [arXiv:hep-th/9912033].











\bibitem{Bern:1998ug}
  Z.~Bern, L.~J.~Dixon, D.~C.~Dunbar, M.~Perelstein and J.~S.~Rozowsky,
  ``On the relationship between Yang-Mills theory and gravity
  and its implication for ultraviolet divergences,''
  Nucl.\ Phys.\ B {\bf 530} (1998) 401
  [hep-th/9802162].


\bibitem{BjerrumBohr:2010ta}
  N.~E.~J.~Bjerrum-Bohr, P.~H.~Damgaard, B.~Feng and T.~Sondergaard,
  ``Gravity and Yang-Mills Amplitude Relations,''
  Phys.\ Rev.\  D {\bf 82}, 107702 (2010)
  [arXiv:1005.4367 [hep-th]].
  N.~E.~J.~Bjerrum-Bohr, P.~H.~Damgaard, B.~Feng and T.~Sondergaard,
  ``New Identities among Gauge Theory Amplitudes,''
  Phys.\ Lett.\ B {\bf 691}, 268 (2010)
  [arXiv:1006.3214 [hep-th]].
  N.~E.~J.~Bjerrum-Bohr, P.~H.~Damgaard, B.~Feng and T.~Sondergaard,
  ``Proof of Gravity and Yang-Mills Amplitude Relations,''
  JHEP {\bf 1009}, 067 (2010)
  [arXiv:1007.3111 [hep-th]].

\bibitem{BjerrumBohr:2010hn}
  N.~E.~J.~Bjerrum-Bohr, P.~H.~Damgaard, T.~Sondergaard and P.~Vanhove,
  ``The Momentum Kernel of Gauge and Gravity Theories,''
  JHEP {\bf 1101} (2011) 001
  [arXiv:1010.3933 [hep-th]].








































\end{thebibliography}
\end{document}